\documentclass[printer]{aa}

\usepackage{caption}
\usepackage{graphicx}
\usepackage[position=top]{subfig}
\usepackage{amsmath}
\usepackage[utf8]{inputenc}
\usepackage{epsfig,amsmath,amssymb}
\usepackage[varg]{txfonts}
\usepackage{academicons}
\usepackage[breaklinks, colorlinks, citecolor=blue]{hyperref}
\usepackage[dvipsnames]{xcolor}
\usepackage{xcolor}
\usepackage{orcidlink}

\usepackage{soul} 
\usepackage{ulem} 

\definecolor{encolor}{HTML}{b3003b}

\definecolor{akcolor}{HTML}{FF007F}

\newcommand{\mb}{\mathbf}

\def\ga{\,\hbox{\hbox{$ > $}\kern -0.8em \lower 1.0ex\hbox{$\sim$}}\,}
\def\la{\,\hbox{\hbox{$ < $}\kern -0.8em \lower 1.0ex\hbox{$\sim$}}\,}
\def\beq{\begin{equation}}
\def\eeq{\end{equation}}

\begin{document}

%
\title{Strong turbulence and magnetic coherent structures in the interstellar medium}
%
\authorrunning{Ntormousi et al.}
\titlerunning{Statistics of magnetic critical points}
\author{Evangelia Ntormousi \orcid{0000-0002-4324-0034}\inst{1,2}, Loukas Vlahos\inst{3}, Anna Konstantinou \orcid{0000-0002-4758-212X} \inst{2,4} and Heinz Isliker\orcid{0000-0001-9782-2294}\inst{3}} 
\date{Received -- / Accepted --}

\institute{Scuola Normale Superiore,
Piazza dei Cavalieri, 7
56126 Pisa, Italy
\\
\and
Institute of Astrophysics,
Foundation for Research and Technology (FORTH), 
Nikolaou Plastira 100, Vassilika Vouton
GR - 711 10, Heraklion, Crete, Greece
\\
\and
Department of Physics, Aristotle University of Thessaloniki, 541 24 Thessaloniki, Greece
\\
\and
Department of Physics and ITCP, University of Crete, 71003 Heraklion, Greece
}

\abstract
{Magnetic turbulence is classified as weak or strong based on the relative amplitude of the magnetic field fluctuations compared to the mean field. These two classifications have different energy transport properties.} 
{The purpose of this study is to analyze turbulence in the interstellar medium (ISM) based on this classification. Specifically, we examine the ISM of simulated galaxies to detect evidence of strong magnetic turbulence and provide statistics on the associated magnetic coherent structures (MCoSs), such as current sheets, that arise in this context.}
{We analyzed MHD galaxy simulations with different initial magnetic field structures (either completely ordered or completely random) and recorded statistics on the magnetic field fluctuations ($\delta B/B_0$) and the MCoSs, which are defined here as regions where the current density surpasses a certain threshold. We also studied the MCoS sizes and kinematics.}
{The magnetic field disturbances in both models follow a log-normal distribution, peaking at values close to unity, which turns into a power-law at large values ($\rm \delta B/B_0 > 1$), consistently with strong magnetic turbulence. The current densities are widely distributed, with non power-law deviations from a log-normal at the largest values.
These deviating values of the current density define MCoSs.
We find that, in both models, MCoSs are fractally distributed in space, with a typical volume-filling factor of about 10 percent, and tend to coincide with peaks of star formation density. Their fractal dimension is close to unity below kpc scales, and between 2 and 3 on larger scales. These values are consistent with MCoSs having a sheet-like or filament-like morphology.}
{Our work challenges the prevailing paradigm of weak magnetic turbulence in the ISM by demonstrating that strong magnetic disturbances occur even when the initial magnetic field is completely ordered. This strong magnetic turbulence arises self-consistently from differential rotation and supernova feedback. Our findings provide a foundation for a strong magnetic turbulence description of the galactic ISM.}


\maketitle

\section{Introduction}
Turbulence is ubiquitous in the interstellar medium (ISM) 
due to the vast Reynolds numbers involved, and can be driven by a variety of processes, such as differential rotation and stellar feedback \citep[e.g.,][]{Elmegreen2004,Bran_laz2013}. 
As a result, the ISM is organized in hierarchical structures that efficiently direct energy  
across a huge range of scales, from kpc to a fraction of a pc.
Magnetic fields are a crucial player in this process: they affect the formation of cold star-forming gas, shape stellar feedback regions \citep[e.g.,][]{pattle_ppvii}, and can also directly influence the turbulent energy cascade \citep[e.g.,][]{Goldreich95,Goldreich_sridhar1997,cho_lazarian_2003}. 
As a consequence, the modern description of ISM dynamics is that of magnetized turbulence \citep[see e.g.,][for a review]{ferriere2020}.

In general, the study of magnetized turbulence can be divided into "weak"  or "wave" turbulence and "strong" turbulence.
Each description makes different assumptions about the magnetic field fluctuations in the plasma, with important implications for the energy cascade and the interaction with charged particles, like cosmic rays.

The "weak" or "wave" turbulence description assumes that the turbulent perturbations are small-amplitude wave packets, the nonlinear interaction of which is slow compared to the wave speed \citep[e.g.,][]{Nazarenko2011, Schekochihin2022}.
This representation of a magnetized plasma is correct if, for example, the fluctuations of the magnetic field 
$\delta \mb{B}$ are very weak compared to the mean ambient field of the plasma, $\mb{B}_0$, or $|\delta \mb{B}|<<|\mb{B}_0|$.
In weak (wave) turbulence, the spectral energy transfer happens through resonant three-wave interaction. This approach is very convenient because it allows the use of quasi-linear theory and prescribed energy dissipation at the small wavelengths \citep{Vedenov63, Galtier09}. 
 
The regime in which unstable waves reach large amplitudes ($|\delta \mb{B}| \geq |\mb{B}_0|$ ) is called "strong" turbulence 
\citep{Goldreich95, Perez08, Schekochihin2022}.   
In this situation,
the non-linear evolution of the magnetic disturbances controls the energy transfer between the different scales and the charged particles. 
The most important characteristic of strong turbulence, which is not present in weak turbulence,  is the intermittent appearance of Magnetic Coherent Structures 
(MCoSs) (Current Sheets (CS), magnetic filaments, large amplitude magnetic disturbances, vortices, and shocklets). MCoSs are collectively the locus of magnetic energy transfer (dissipation) into particle kinetic energy, leading to heating and/or acceleration of the latter \citep[see more details in][]{Vlahos23}.  

Weak and strong turbulence are not mutually exclusive situations. For example, small disturbances on large scales can grow nonlinear on small scales, so the weak-to-strong turbulence transition is scale-dependent \citep{Goldreich95,Goldreich_sridhar1997,Schekochihin2012,Meyrand2018,Fornieri2021}. 
One way to describe this process is through the anisotropy of the turbulence when the plasma is threaded by a large-scale mean field. \citet{Goldreich95, Goldreich_sridhar1997} did so by introducing an anisotropy parameter, defined as the product of two ratios: the ratio of the wavenumbers parallel and perpendicular to the local mean magnetic field, and the ratio of the speed at a certain scale to the Alfv\'{e}n speed.

However, in many space and astrophysical plasmas the magnetic fluctuations are very localized, occupying small volumes of otherwise largely isotropic systems. There, the scale dependence of the anisotropy is not a good metric for characterizing turbulence as weak or strong. For this reason, more recent literature, particularly the part dealing with CR transport, adopts a wider definition of strong turbulence, namely a plasma with $\delta B/B_0 \geq 1$ \citep[e.g.,][]{Lemoine2023, Kempski23, Butsky2024}. This is also the definition we will adopt in this work.

In strongly turbulent, magnetized plasmas, 
MCoSs, and especially CSs, 
are also evolving and fragmenting, becoming locally the source of new clusters of MCoSs. 
It has been shown that MCoSs follow monofractal or multifractal scalings, both in space and laboratory plasmas \citep{Tu95,Shivamoggi97,Biskamp03,Dimitropoulou13,Leonardis13,Schaffner15,Isliker19, Consolini23}. In particular, the CSs inside a turbulent reconnection volume are fractally distributed in space \citep{Vlahos04}.

In general, strong magnetic turbulence can be generated by (i) the nonlinear coupling of large amplitude unstable plasma modes, (ii) by the explosive reorganization of large scale magnetic fields, or (iii) by the fragmentation of existing MCoSs. 

From theoretical studies of the magnetized ISM, we know that at least situations (i) at (ii) occur frequently in galaxies. For example, large-amplitude, unstable modes can be created by magnetic buoyancy instabilities \citep{Parker1966,Mouschovias1974} or the Magneto-Rotational Instability (MRI) \citep[][]{korpi_maclow2003,Kitchatinov2004}. Rapid, large-scale magnetic field reorganization can happen in regions of strong shear \citep[e.g.,][]{fraser2021,Tripathi2023}, and is required during the growth phase of the galactic magnetic field through a dynamo \citep{Bran_sub2005,Bran_EN23}.

Still, so far no study has studied the emergence of strong magnetic turbulence in the ISM, and whether situation (iii) also arises in the ISM is an unexplored subject.

Eventually, the strength and structure of the turbulence and the MCoSs will depend on many factors, such as the properties of the gas, the geometry and strength of the large-scale magnetic field, 
the sources and sinks of energy (which for the ISM can be, for example, stellar feedback and cooling, respectively), and the boundary conditions 
\citep{Subramanian06, Kivodides07, Kritsuk17, Richard22, Colman22, Galishnikova2022, Lubke24, Lesaffre24}.
Therefore, the characterization of interstellar turbulence in terms of its strength in different situations is of great interest for understanding the energy transfer process in the ISM. 

In this work, we look for characteristics of strong magnetic turbulence in MHD simulations of galaxies. Specifically, we study the distribution of magnetic disturbances, identify MCoSs, and describe their statistics. 


We describe the numerical simulations and the MCoS identification methods in Section \ref{sec:meth} and present the results in Section \ref{sec:res}. We discuss our findings in Section \ref{sec:dis} and draw our conclusions in Section \ref{sec:concl}. 

\section{Methods}
\label{sec:meth}

\subsection{Numerical simulations}

As a test case for the self-consistent generation of strong magnetic turbulence, we use two numerical simulations of galaxy evolution, presented in \citet{Konstantinou2024}. 
These multi-physics simulations follow the evolution of Milky-Way-like galaxies under different initial conditions for the magnetic field, which makes them ideal for studying any resulting differences in the statistics of MCoSs. Here we describe the main features of these models, but we refer the reader to the presentation 
paper for more details.

\subsubsection{Code and initial conditions}

The simulations were performed with the MHD, Adaptive Mesh Refinement (AMR) code RAMSES \citep{teyssier2002,fromang2006}. RAMSES treats collisionless components like dark matter and stars with a Particle Mesh (PM) method and the MHD fluid on a Cartesian AMR grid with a Godunov method. The magnetic field is evolved on a staggered mesh to fulfill the solenoidal condition.

For these simulations, \citet{Konstantinou2024} use a customized version of RAMSES that follows the non-equilibrium chemistry of $H_2$ formation and dissociation through the KROME package (see \citealt{grassi2014} for the KROME package, \citealt{pallottini2017} and \citealt{decataldo2020} for previous applications). This implementation allows for a more physical description on the star-formation process, based on the molecular hydrogen ($H_2$) content of the cells.

The models represent local Milky-Way-sized galaxies, with a virial velocity of 200 km/s, which corresponds to a mass of $10^{12} M_{\odot}$. The dark matter (DM) halo makes up 97.5$\%$ of the total mass, the thin stellar disk 1.425$\%$, the gaseous disk 0.075$\%$, and the gaseous halo 1$\%$.
The DM and gaseous haloes follow a pseudo-isothermal density profile with a scale length of 3 kpc and a radial density cut at 50 kpc. The thin stellar disk follows a Myamoto-Nagai profile \citep{miyamoto} with a scale length of 3 kpc and a radial density cut at 12 kpc, and the gaseous disk follows an exponential density model with a scale length of 4 kpc and a radial density cut at 15 kpc. The initial temperature of the gas is 8000 K, without any initial turbulent velocity.

The models are placed in a cubic box of 100~kpc size, with periodic boundary conditions. The coarsest level of refinement corresponds to $128^3$ cells, and the highest to $4096^3$ cells. The refinement for levels up to $512^3$ is geometry-based, in cylindrical regions of decreasing size centered on the galaxy that fully encompass the gaseous disk at all times. 
Three additional levels of refinement, up to $4096^3$, are triggered by a Jeans-based criterion, namely requiring that the local Jeans length is resolved with at least 10 cells. The maximum resolution corresponds to 24 pc.

These initial conditions are identical for the two simulations, but their initial magnetic fields are different. 
Specifically, one simulation (hereafter model T) starts with a toroidal magnetic field and the other with a random field (hereafter model R). The magnetic field of model R has a power spectrum with a $\rm k^{-3}$ scale dependence.
Both fields have a maximum value of $\rm 10~\mu G$ at the center of the galaxy, exponentially dropping with radius and height. The scale height and length of the magnetic field are both set to 1~kpc. This difference in the initial conditions means that any emerging randomness in model T's magnetic field, or 
order in model R's magnetic field, as the models evolve, will be due to the nonlinear evolution of the galaxy's interstellar medium.

\subsubsection{Turbulence in the simulations}

Since the initial conditions contain no turbulent velocities, two mechanisms create turbulent flows as the models evolve: stellar feedback and differential rotation. These processes are identically modeled in the two simulations. 

Specifically, both models include star formation based on a threshold in the $H_2$ content of the cells. They also include the same supernova feedback recipe, where thermal energy is injected in the 27 cells around the exploding stellar particle. While the supernova feedback and the differential rotation of the galaxy self-consistently drive turbulent velocities in both models, we are interested in the emergence of strong fluctuations in the magnetic field, and the consequent appearance of MCoSs.

Model T initially only has an ordered magnetic field, and model R only a random one. Therefore, after the turbulence has reached a steady state in the two models, any differences or similarities between them in terms of the magnetic field perturbations can inform us on the mechanisms that create MCoSs. In the following we describe our definitions for strong magnetic turbulence and MCoSs and present their statistics at different times in the model evolution.

\subsection{Defining strong magnetic turbulence}

Since we are interested in MCoSs appearing in strong magnetic turbulence, we first identify the regions where the magnetic field fluctuations are at least comparable to the mean field, $\rm |\delta \mb{B}|\geq |\mb{B}_0|$.

Defining $\rm \mb{B}_0$ is not straightforward because different choices for the averaging can give different answers \citep[e.g.,][]{Gent2013,Bran_EN23}. 
Here, in order to be able to define $\rm {\bf{B_0}}$ 
both in a predominantly random and a predominantly ordered environment, we employ local filtering of the magnetic field. 

Specifically, for each cell $i$, we find the $N$ closest neighbors by constructing a K-D Tree \citep{Maneewongvatana_mount1999}, choosing a fixed value for N.
We then use the $N+1$ values of the magnetic field to define $\rm{\bf{B_{0}}}$ through the local median value of each component, $\rm B_{0}\hat{x},~B_{0}\hat{y},~B_{0}\hat{z}$ and $\rm\delta{\bf{B(i)}} = {\bf{B_{tot}}}(i)-{\bf{B_{0}}}(i)$.
This method is a significant improvement on the definition used in \citet{EN2020}, who interpolated the AMR data onto a regular grid and then applied a median kernel to obtain $\rm {\bf{B_0}}$, because here we make use of the full AMR information. 
We note that we also tried azimuthal averaging for defining $\rm {\bf{B_{0}}}$ and we found that it produced unwanted artifacts at the radial 
edges of the azimuthal bins.

Since the gas density and the magnetic field outside the galactic disk remain negligible throughout the simulation time (which means that also the grid resolution is poor in these areas), the above process is applied to the disk regions only. This includes a cylinder of radius $\rm R_{g}=14~kpc$ and height $\rm |z_g|=3~kpc$ centered on the galactic center. We have confirmed that the results do not change if we include larger portions of the galaxy, apart from the inclusion of noise.

\begin{figure*}[h!]
    \centering
   \includegraphics[trim={0.5cm 1.5cm 2cm 0cm},clip,width=0.8\textwidth]{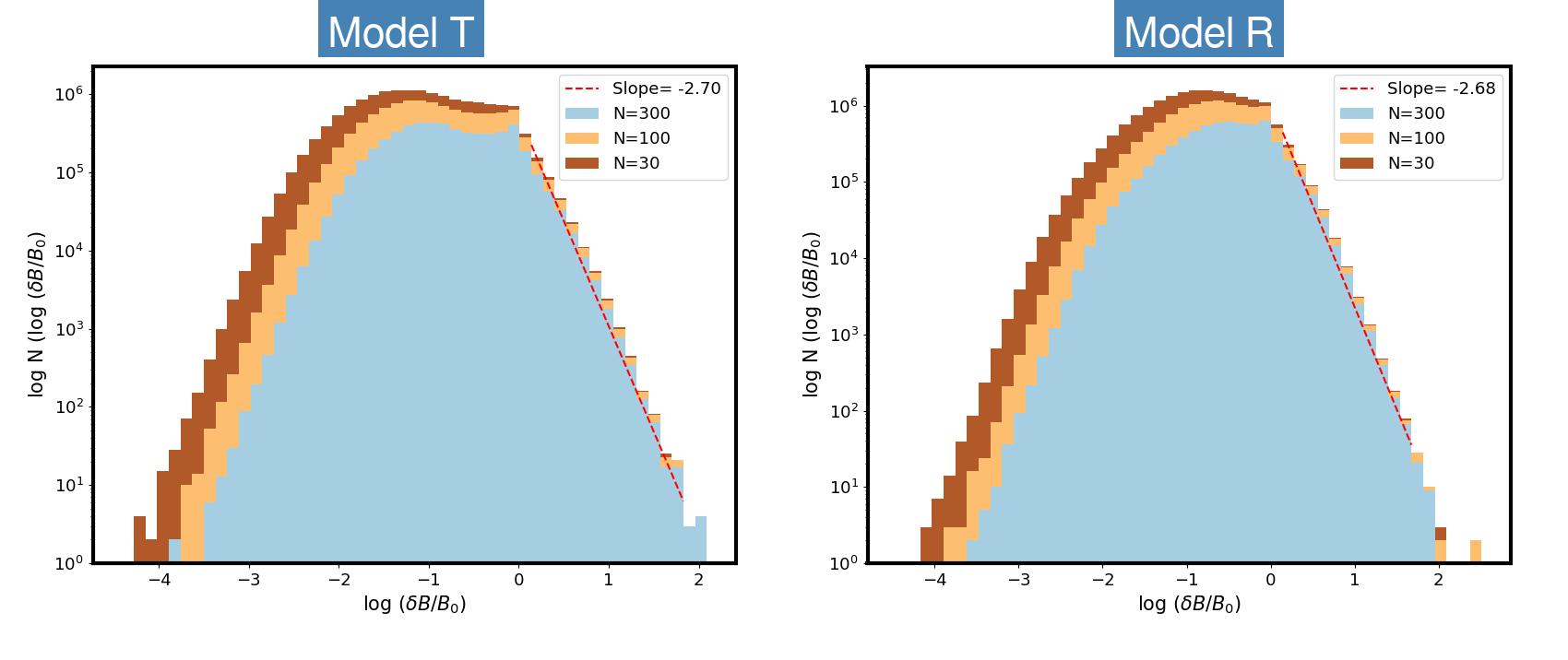}
    \caption{PDFs of $\rm log~\delta B/B_0$, where $\rm B_0$ is the median-filtered magnetic field over a set of N neighbors, at time t=500~Myrs. The dashed red line is a fit to the power-law slope above $\rm log~\delta B/B_0= 0$.}
   \label{fig:pdfs_deltab_kernel}
\end{figure*}

In Fig. \ref{fig:pdfs_deltab_kernel} we plot the probability density functions (PDFs) of $\rm log~\delta B/B_0$, where $\rm \delta B$ and $\rm B_0$ are the magnitudes of the vectors $\rm \delta{\bf{B}}$ and $\rm {\bf{B_0}}$, for different values of N for the two models at the same evolution time (500~Myrs). 
We immediately notice that the peak value of $\rm \left(\delta B/B_0\right)_{peak}\simeq 1.2$ is almost the same for both models, as is the maximum $\rm\left(\delta B/B_0\right)_{max}\simeq100$.
(The peak value in model R is actually at $\rm \delta B/B_0\simeq 2$, slightly higher than $\rm \delta B/B_0\simeq 1.17$ in model T).
Also, very interestingly, both sets of PDFs are log-normal up to the peak value, with a power-law tail above $\rm\delta B/B_0>1$, indicative of a dynamical process at play. The power-law slope at high values of $\rm log \left(\delta B/B_0\right)$ is also identical, and equal to about $\alpha=-2.7$.
This similarity between the two sets of distributions is particularly intriguing, given the drastically different initial conditions ($\rm \delta B$ is initially zero for model T, and $B_0$ is initially zero
for model R). Even more surprising is that this similarity, including the power-law slope, is established very early on in the simulations (see Appendix \ref{app:J_Jrms}).

The effect of changing N is quite pronounced for the low-$\rm\delta B/B_0$ end of the distribution, but the peak and the high-$\rm\delta B/B_0$ behavior are largely unaffected. 
Since we are interested in this dynamical behavior at large $\rm\delta B/B_0$, where the PDFs do not depend on N, in the following sections we will present results for the intermediate value N=100.

\subsection{Defining magnetic coherent structures}
\label{sec:defmcos}

Identifying MCoSs in numerical simulations of turbulence is notoriously hard \citep{Vlahos23}. 

Here, we define MCoSs as regions of strong current density \citep[e.g.,][]{Uritsky2010,Zhdankin13, Sisti21}. 
Regions with current density $\rm J\equiv\bf{|J|}$ larger than the rms average $\rm J_{rms}\equiv\bf{|J_{rms|}}$ will be sites of strong energy dissipation, where particles are expected to undergo violent changes in their trajectory. Since the current is not part of the code's output, we calculate the current density from the simulations as $\rm J=\nabla\times\bf{B}$ using each cell's closest neighbors from the AMR's octree structure for the spatial derivatives.

\begin{figure*}
    \centering
 \includegraphics[trim={0.5cm 1.5cm 2cm 0cm},clip,width=0.8\textwidth]{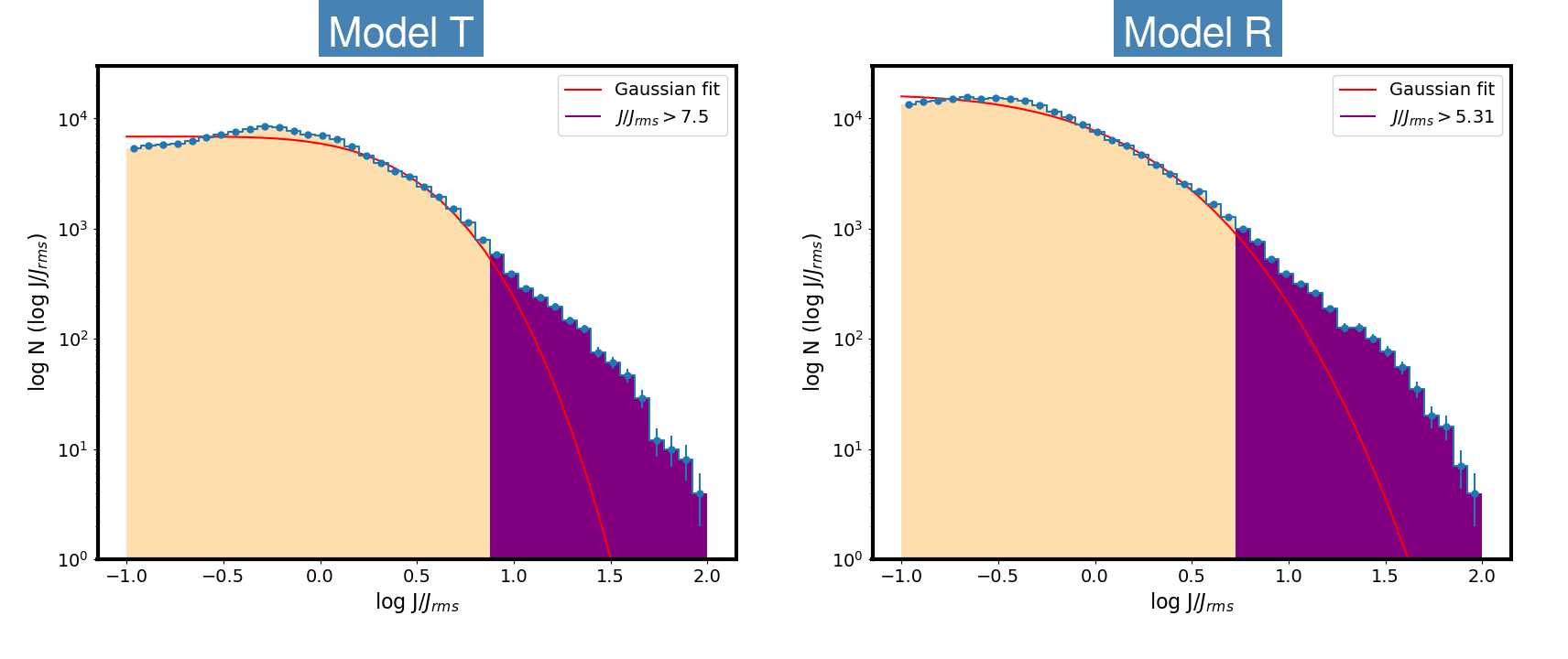}
    \caption{PDFs of the logarithm of the current density normalized by its rms value, $\rm log~J/J_{rms}$ for the two models at t=500~Myrs. For clarity, here we only show the high-value end of the distribution, that extends down to near-zero values. The red curve shows a Gaussian fit to the distribution.}
    \label{fig:jdistributions}
\end{figure*}

Figure \ref{fig:jdistributions} shows the distribution of the logarithm of $\rm J/J_{rms}$ in the two models at 500~Myrs of evolution. Instead of explicitly limiting the range of the plot to the disk, like for the $\rm \delta B/B_0$ distributions, here we bin only the highest $\rm J/J_{rms}$ values.
Not surprisingly, all these locations actually do reside in the galactic disk. The actual range of $\rm J/J_{rms}$ values in each snapshot spans several orders of magnitude, with many locations containing current density close to zero. Similar plots for other snapshots are included in Appendix \ref{app:J_Jrms}. 
Like $\rm \delta B/B_0$, also $\rm J/J_{rms}$ deviates from a log-normal distribution at the highest values, although in this case there is no clear power-law behavior. 
In Fig. \ref{fig:jdistributions} we have marked the point where the deviation from a log-normal begins, which we will call $\rm \left(J/J_{rms}\right)_{c}$, with purple color. In the following, we will use this value as a threshold for defining MCoSs. In other words, a MCoS in our definition will be defined by the locations where $\rm J/J_{rms} >\left(J/J_{rms}\right)_{c}$.
Unlike the change of behavior of the $\rm \delta B/B_0$ distribution, which happens consistently at unity for all snapshots and both models, the value of $\rm \left(J/J_{rms}\right)_{c}$ differs between snapshots and models.

\subsection{Measuring the fractal dimension of MCoSs}

Stochastic or chaotic systems often present some degree of self-similarity up to a natural scale where the dominant dynamical process changes. A good example of this behavior is a turbulent cascade, where self-similarity characterizes the system from the integral to the diffusive scale. A natural metric to describe this property is the fractal dimension of a measurable quantity, such as, in the example of turbulence, the density distribution of strong fluctuations. In this context, the fractal dimension can be interpreted as a surface- or volume-filling factor of the observed structures.

In the ISM, the fractal dimension has been widely used to characterize the complex dynamics of turbulence and gravity in molecular clouds \citep[e.g.,][]{Bazell1988,Elmegreen_Falgarone1996,Stanimirovic1999,Sanchez2005}. Here we will use the fractal dimension to characterize the distribution of MCoSs. 

For this calculation, we will use the simplest approach, which is the box counting technique \citep[e.g.][]{Falconer1990}. For each output, our method calculates the box counting dimension of the locations with $\rm J/J_{rms} >\left(J/J_{rms}\right)_{c}$, with $\rm \left(J/J_{rms}\right)_{c}$ as defined in Sec. \ref{sec:defmcos}.  

\section{Results}
\label{sec:res}

\subsection{Spatial distribution of MCoSs}
\label{sec:mcos_distribution}
As a first step we want to study the spatial distribution of MCoSs in the simulations and see if there is any correlation with other quantities, such as the local magnetic field fluctuations, density, and star formation rate, that could provide intuition into the physical processes that drive their formation. 

\begin{figure*}
 \includegraphics[width=0.97\textwidth] {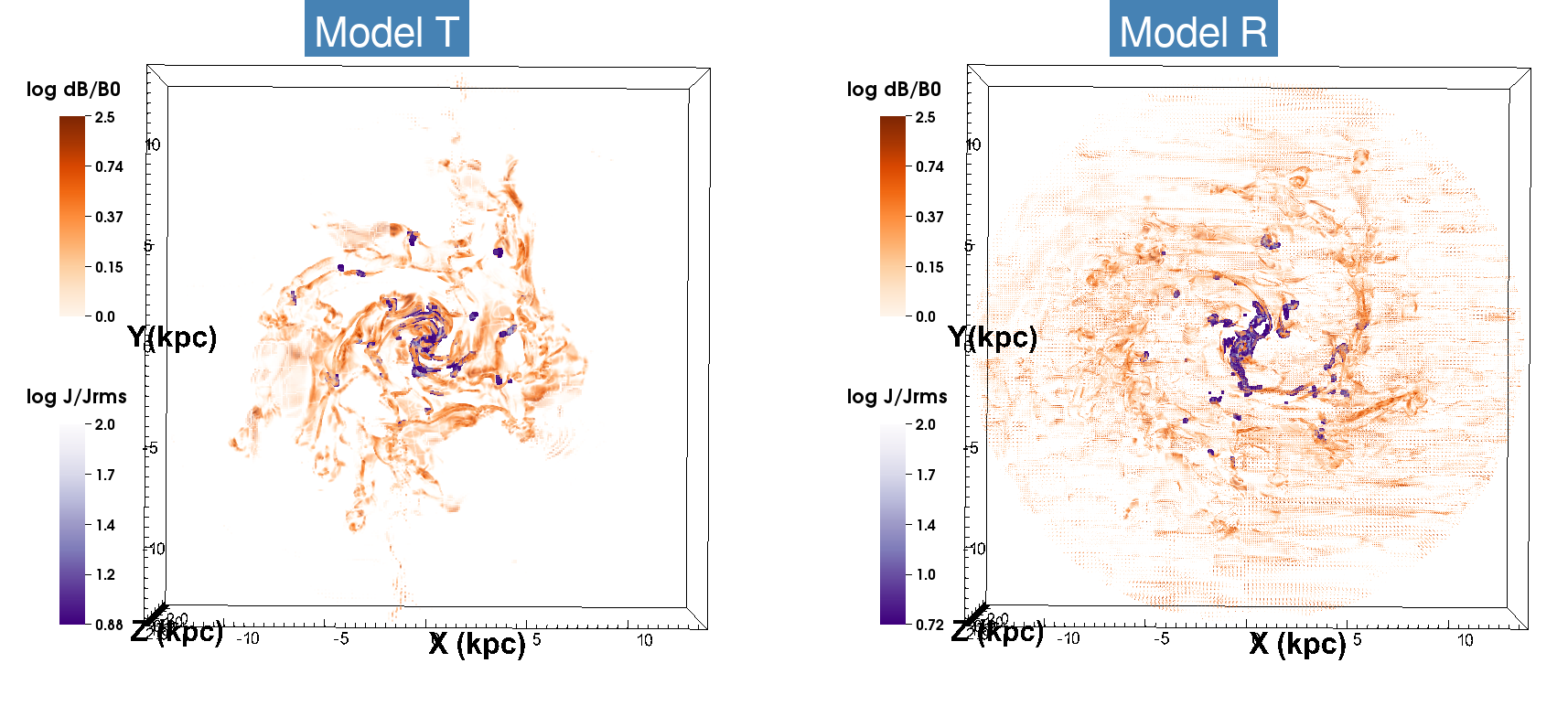} \\ 
\includegraphics[width=0.95\textwidth,trim={0cm 0cm 0cm 0cm}]{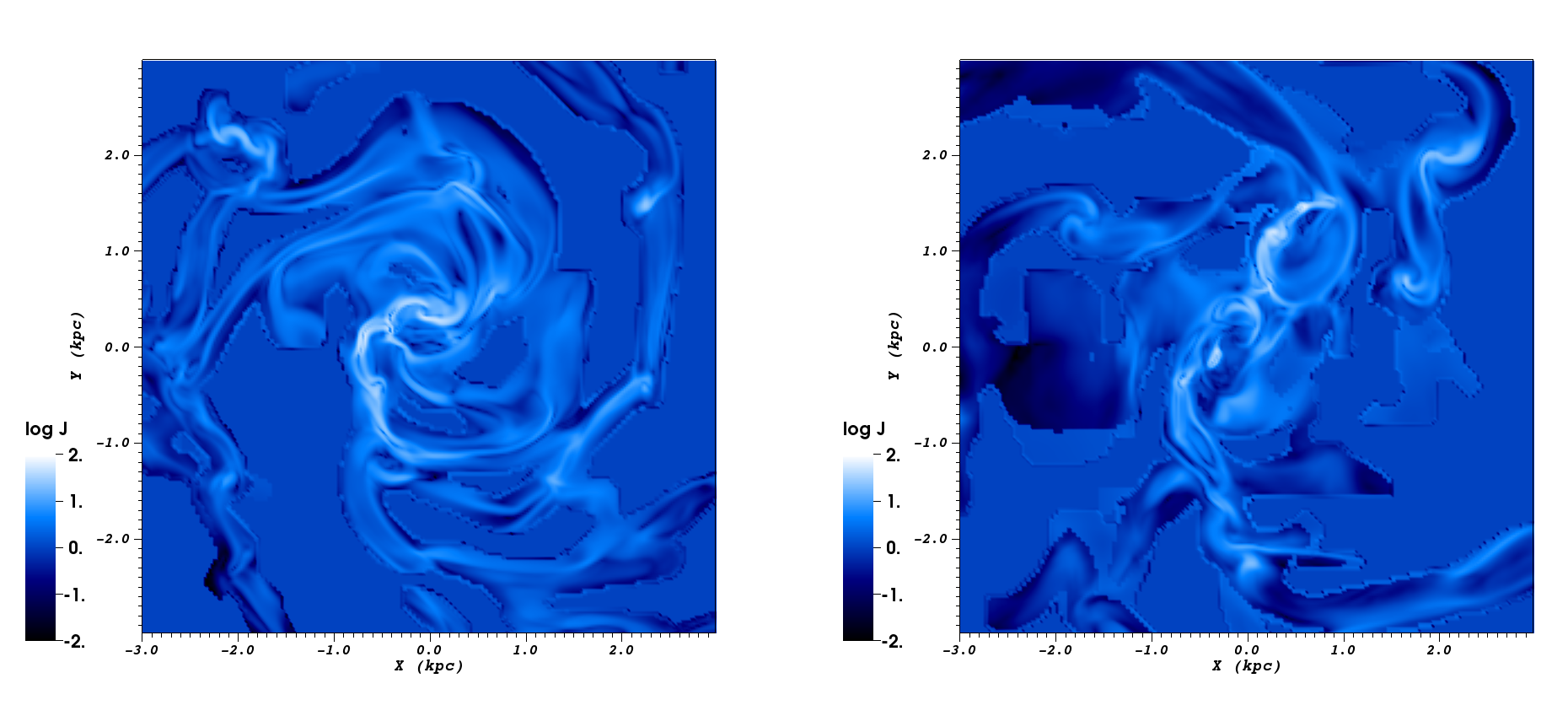} 
    \caption{Top panel: Volume rendering of $\rm \delta B/B_0$ (orange), shown only for values above unity, and current density normalized by its rms value (purple), shown only for values above the critical for each model. Bottom panel: Zoom-in in the central 6~kpc of each run, showing a midplane cut of the current density, in code units.
    Here the colorbar range has been limited to the highest J values. 
    The time corresponding to these snapshots is 500 Myrs.}
    \label{fig:db_b0_crr_volume}
\end{figure*}

First, we focus on the local magnetic field fluctuations, measured by $\rm \delta B/B_0$. In the top panel of Fig. \ref{fig:db_b0_crr_volume}, we plot the 3D distribution of $\rm \delta B/B_0$ together with that of $\rm J/J_{rms}$ for the 500~Myr snapshot of the two models. Interestingly, regions of high $\rm J/J_{rms}$ are contained within regions of high $\rm\delta B/B_0$, a correlation consistent with a picture in which strong magnetic turbulence activates regions of high electric current. 
However, for both models, regions of high $\rm J/J_{rms}$ seem to concentrate near the galactic center, while regions of high $\rm\delta B/B_0$ extend to larger radii. Since by construction, the models have a higher density and magnetic field in the central regions, this spatial distribution could indicate a more direct connection between $\rm J/J_{rms}$ and these quantities.

In the bottom panel of the same figure we show a zoomed-in view of the midplane current density (in code units), which highlights the filamentary structure of the currents, and confirms their classification as CSs. While the two models follow a similar evolution in terms of their star formation rate and overall balance between magnetic and thermal energy \citep[as pointed out by][]{Konstantinou2024}, the morphology of the currents in the central regions is visually very different, possibly due to the complex nature of the system, that introduces stochasticity. 

\begin{figure*}
    \centering
 \includegraphics[width=\textwidth,trim={0 0cm 8cm 0},clip]{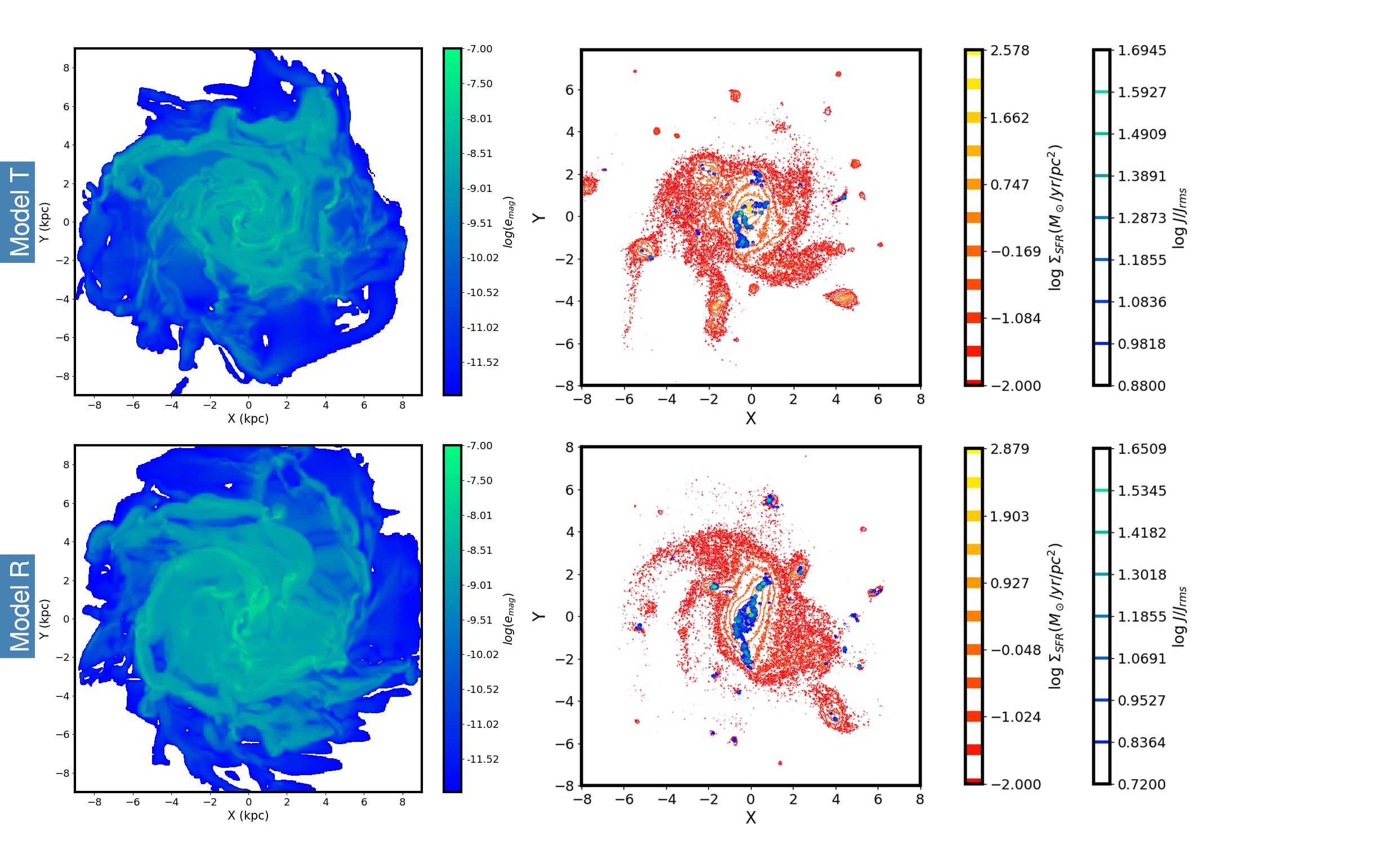}
    \caption{Midplane slices of magnetic energy (left) and current density, shown here only for values above the critical for each model, overlaid with star formation rate surface density (right) for models T and R (toroidal and random initial magnetic field, respectively), at 500 Myrs of evolution.}
    \label{fig:mag_current_slices}
\end{figure*}

Throughout the simulation, and despite the different initial conditions in the two models, the regions of high magnetic energy and current density are concentrated around the galactic spiral arms. Fig. \ref{fig:mag_current_slices} provides an illustration. There we plot the magnetic energy and current density at time t=500~Myrs, for the two models. 
Interestingly, we find a strong spatial correlation between the current density and the star formation rate surface density, shown in Fig. \ref{fig:mag_current_slices} as contour lines on the $\rm J/J_{rms}$ maps. This correlation, which essentially is a connection between the local gas density peaks and the current density, is more pronounced than that between the magnetic field strength and the current density. We also studied the connection between current density and vorticity, and between current density and $\nabla\cdot \mb{v}$ (not shown here), finding a spatial correlation weaker than that between magnetic field strength and current density. This finding implies that the MCoS in these models are closely connected to the process of gravitational collapse, which strongly tangles magnetic field lines.

\subsection{Properties of MCoS}

Since MCoSs are good candidates for energy dissipation and particle acceleration,  we are interested in their typical size and volume-filling factor. We define MCoSs as clusters of locations where $\rm J/J_{rms}$ is above a threshold $\rm \left(J/J_{rms}\right)_{c}$
, where the threshold is identified in the $\rm J/J_{rms}$ PDFs as in Fig. \ref{fig:jdistributions}. If two cells i and j have $\rm J/J_{rms}$ above the threshold and are separated by less than $\rm dx_i+dx_j$, where $\rm dx_k$ is the edge size of cell k, then they are considered part of the same cluster.

\begin{figure*}
    \centering
    \includegraphics[trim={2.5cm 1.5cm 1.5cm 0cm},clip,width=0.8\textwidth]{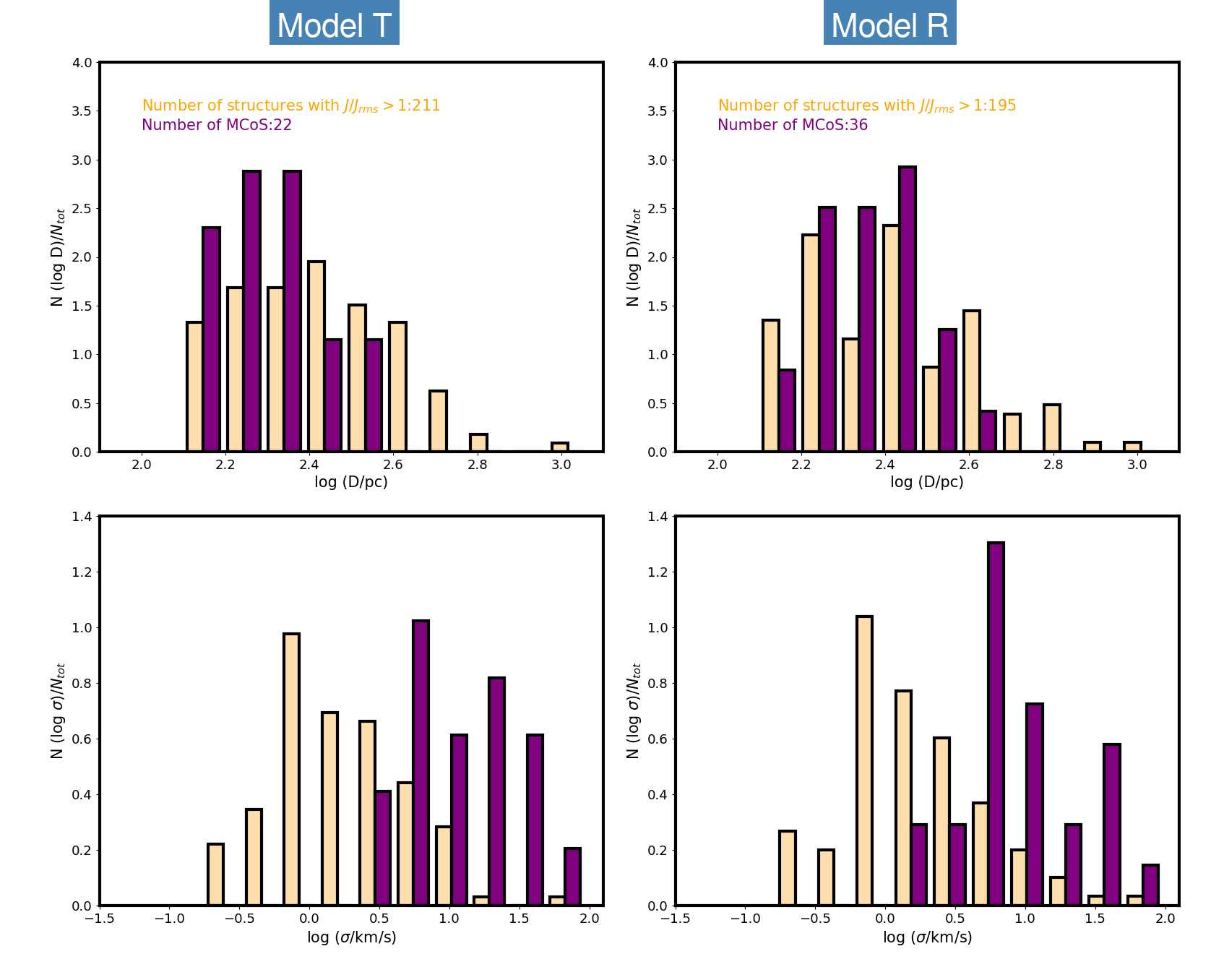}
    \caption{Histograms of MCoS sizes (top) and internal velocity dispersions (bottom) in the 500~Myr snapshot of the two models. The light orange histograms correspond to clusters of locations with $\rm J/J_{rms}>1$, while the purple ones to clusters of locations with $\rm J/J_{rms}>\left(J/J_{rms}\right)_c$, which we identify as MCoSs.}
    \label{fig:mcos_size_sigma_histograms}
\end{figure*}

The MCoS size PDFs are shown in the top panel of Fig. \ref{fig:mcos_size_sigma_histograms} in purple color. Here, size is calculated as the cube root of the structure volume. In each panel of the figure we have also included distributions where structures are identified among all locations with $\rm J/J_{rms} > 1$, shown in light orange. All distributions contain only cell clusters with more than 16 members and are normalized by total number of counts and bin width. 
We notice that the size PDFs of MCoSs are quite narrow, ranging from about 125 to about 400~pc, and peaking around 200~pc. Including all locations where $\rm J/J_{rms} > 1$ does not significantly widen the size range.  It is worth noting that this typical MCoS size is of the order of the disk scale height, or the size of a typical dense cloud region. The latter connection is particularly interesting, since we also noticed a correlation between $\rm J/J_{rms}$ ans $\Sigma_{SFR}$ (see Section \ref{sec:mcos_distribution}).
The total number of MCoSs identified in each snapshot is quite low: 22 in model T, and 36 in model R, and including locations with lower $\rm J/J_{rms} > 1$ does not significantly increase the number of structures: 211 in model T versus 195 in model R. However, we should note that these simulations are most likely not converged in terms of the typical structure number and size. 

The systemic velocities of all MCoSs, in both models (not shown here), are within one standard deviation from the mean galactic gas motion at their location, meaning that they can move towards each other and merge as the galactic flow evolves, but they are not moving at a peculiar speed.  
Their internal velocity dispersions, shown in the bottom panel of Fig.\ \ref{fig:mcos_size_sigma_histograms}, peak at roughly 6km/sec in both models, reaching up to 56~km/s for a handful of structures. These velocity dispersions are typical of molecular cloud regions in these simulations and in the galactic ISM.

\begin{figure}
    \centering
 \includegraphics[width=\linewidth,trim={2.5cm 1.5cm 2.5cm 1.5cm},clip]{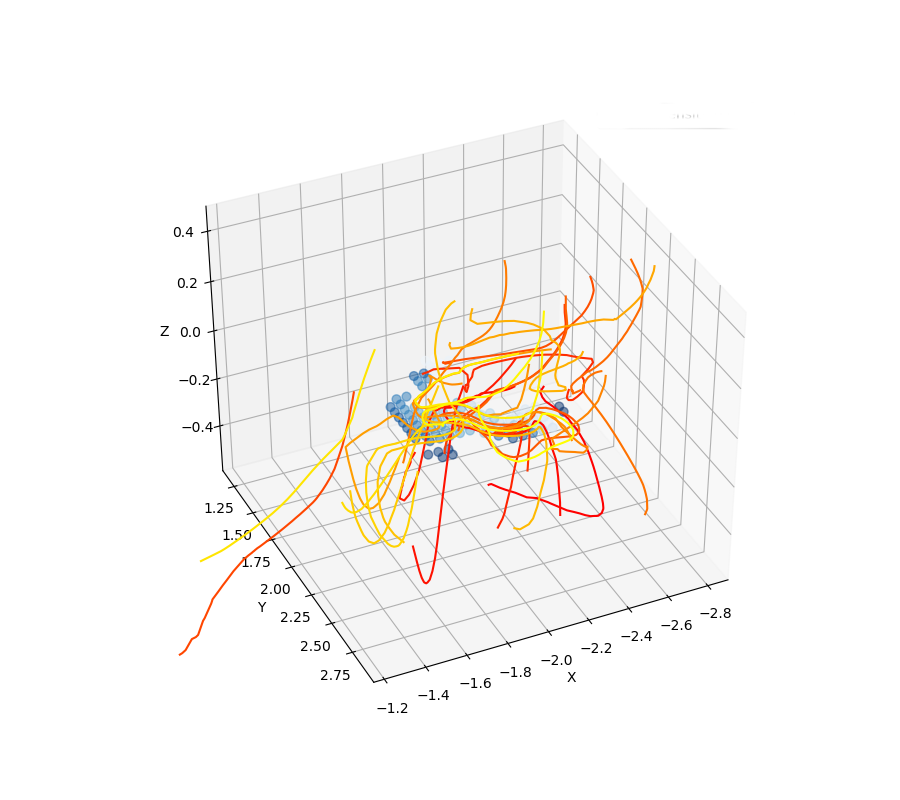}
    \caption{
    Example of a MCoS, taken from a snapshot of model T at 500 Myrs. Magnetic field lines are shown in yellow and orange colors. The blue dots show the current density of each location in code units, with darker colors indicating higher intensity.
    The magnetic field around MCoSs in both models has similarly complex structure.
    }
    \label{fig:mag_field_lines}
\end{figure}

Recent studies using MHD simulations of driven turbulence in a box \citep{Lemoine21, Kempski23}, have identified sharp bends in the magnetic field as a crucial ingredient in CR transport. In this study, we systematically see such features around MCoSs.
Fig. \ref{fig:mag_field_lines} shows an example, with the current density plotted in code units and randomly selected magnetic field lines drawn around it. Similarly to all the MCoSs in the simulations, it hosts a magnetic field with loops and strong bends. This complex structure of the magnetic field suggests that MCoSs are a fundamental feature for studying CR transport in the ISM.

Further, we notice that MCoSs are generally far from spherical, showing a variety of elongated shapes. 
The velocity field around the structures (where the systemic velocity of the structure has been removed) also shows a variety of morphologies: in some cases, there is a clear vorticity pattern, like the structures at the top-left and bottom-right corner of the figure, while in others we can see compession, like the structure at the top-right and bottom-right corner of the figure. However, the velocity field is usually quite complex. This complexity explains the weak correlation between MCoSs and vorticity or $\nabla\cdot \mb{v}$ mentioned in the previous section.

\begin{figure*}[h!]
    \centering
    \includegraphics[width=0.8\textwidth,trim={0 1.5cm 0 0},clip]{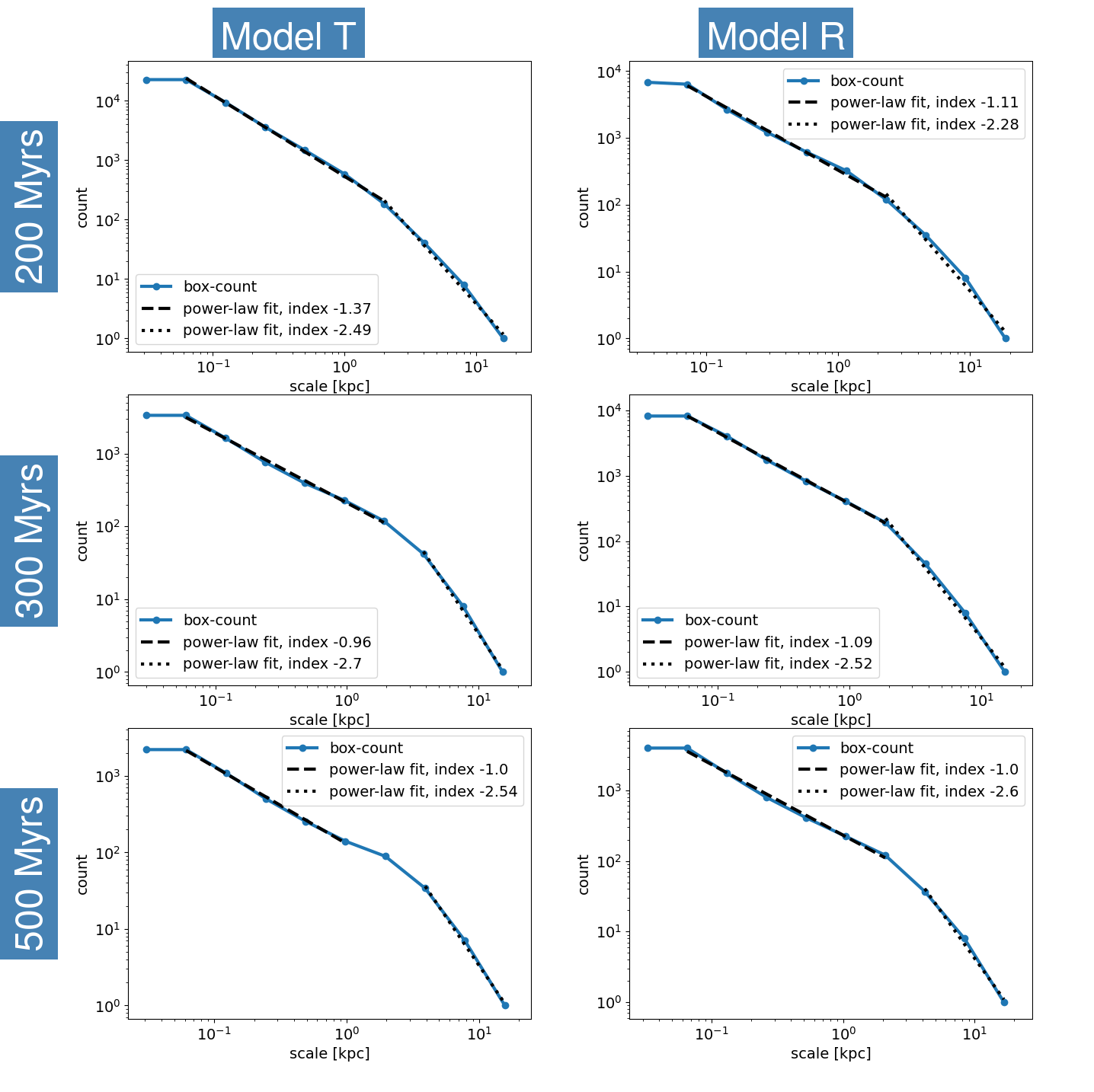}
    \caption{The fractal dimension of structures with $\rm\delta J/J_{rms}$ above a given threshold value, at different times.}
    \label{fig:fractal_dim_time}
\end{figure*}

Very useful information is also contained in the volume-filling factor of MCoSs as a function of scale, as measured by their fractal dimension. In Fig. \ref{fig:fractal_dim_time} we show the box-counting fractal dimension calculation at different simulation times. Interestingly, at all snapshots and for both models, the box counting yields two distinct scalings, corresponding to a fractal dimension:
One scaling is identified at scales below 1~kpc, and is consistently close to 1. This scaling, which can be interpreted as MCoSs resembling filaments or ribbons, is consistent with the elongated structure we identified through the clustering process. 
The second fractal dimension, at scales above 1~kpc, is around 2.5, which can be interpreted as "porous" 3D structures. 
It is worth mentioning, however, that the fractal dimension is quite sensitive to the selected threshold: 
Increasing the threshold gradually selects smaller portions of the regions, naturally leading to a smaller fractal dimension, while the opposite is true when lowering the threshold.
The threshold we selected here is physically motivated, since it signifies the non-Gaussian limit of the normalized current density distribution, a natural indicator of a change of behavior. However, this finding should still be validated with higher-resolution simulations. 

\section{Discussion}
\label{sec:dis}

Even though a plethora of numerical simulations have shown the ubiquity of high Mach number flows in galaxies, to our knowledge, this is the first numerical study to focus on the distribution of magnetic field fluctuations on galactic scales. Specifically, here we focused on determining whether the magnetic turbulence in the ISM should be considered strong or weak by studying the magnetic field fluctuations and the development of MCoSs in MHD numerical simulations of isolated spiral galaxies. 

We found that in models of massive, gas-poor, quiescent and strongly magnetized galaxies, 
the magnetic field fluctuations 
range from $10^{-4}$ to about 100 times higher than the mean value of the mean field, independently of the initial morphology of the magnetic field. Moreover, the interstellar turbulence in these models consistently produces a power-law distribution of $\rm \delta B/B_0$ in the regime above unity. 
This finding shows that strong magnetic turbulence co-exists with weak turbulence in the galactic ISM, covering a significant part of the galactic disk. Specifically, in our models, the strongest fluctuations are clearly concentrated in and around spiral arms.

The ubiquity of strong magnetic turbulence in the disk is consistent with the existing observational probes of magnetic fields in spiral galaxies. For example, \citet{gaensler2011} found strong variations in the Milky Way's Faraday rotation signal, organized in filament-like structures. These structures in Faraday rotation can either be associated with strong variations in the local magnetic field or electron density. Later, in characterizing the Galactic magnetic field in dust polarization, \citet{PlanckXLIV} were able to fit the observed data with a $\delta B/B_0$ of about 0.9.
There is observational evidence of strong magnetic fluctuations also in external galaxies. For instance, using HAWC+ dust polarization data for the M51 galaxy, \citet{Borlaff2021} showed that the magnetic field around the spiral arms is more strongly fluctuating than in the interarm regions. The spatial distribution of $\rm \delta B/B_0$ in our analysis (Fig. \ref{fig:db_b0_crr_volume}) is consistent with this picture. According to similar analyses in the starburst galaxy M82 \citep{LR2021} and in the Antennae merging system \citep{LR2023}, we can expect $\rm \delta B/B_0$ to be even stronger in these more dynamical environments than what we estimated here. Clearly, a wider range of simulation data is needed to predict the $\rm \delta B/B_0$ statistics in more dynamical ISM conditions.

Finally, the simulations we used in this analysis do not host a dynamo. In a turbulent dynamo situation, we also expect the $\rm \delta B/B_0$ distribution to extend to even higher values \citep[See, e.g. recent numerical simulations by][]{gent2024}.

We identify MCoSs in the simulations by looking for regions of high current density, which is equivalent to studying regions of strong dissipation. 
In general, strong dissipation in MHD turbulence is organized in thin sheets. \cite{Lehmann16} with the SHOCKFIND algorithm and \cite{Richard22} designed several criteria that allow for the careful selection of strong dissipation regions and characterize their physical nature as fast shocks, slow shocks, rotational discontinuities or Parker sheets. Our analysis does not go into the details of further categorizing the dissipation regions, but confirms their sheet-like morphology, showing strong currents organized in thin, elongated structures.

The fractal dimension of the MCoSs, which is between 2 and 3 at scales above 1~kpc and close to 1 at scales below 1~kpc, is also consistent with sheet-like structures containing smaller-scale filaments. Due to the limited resolution of these simulations, we are not able to witness any smaller-scale fragmentation of these MCoSs, so follow-up work at higher resolution is needed to elucidate this point. Eventually, at sufficiently high resolution, the final state of this MCoS network should be a turbulent reconnection environment \citep[see][for articles and reviews on this topic]{Matthaeus11, Lazarian12, Karimabadi13a, Karimabadi03b, Karibabadi2013c,Karimabadi2014, Vlahos23}.

The typical number of MCoSs at a given snapshot is of the order of a few tens, rendering a number density of $10^{-7}$ structures per $\rm pc^{-3}$. With the typical structure size of the order of 100~pc, this means that the MCoSs cover about 10 percent of the 
galactic disk.
However, we should stress that, due to the limited resolution of our models in the halo, we could not calculate MCoS statistics in the entire volume of the galaxy. In future work we will expand our study to include the circum-galactic medium.

These findings are of particular relevance for CR propagation models.
Many studies in the past few years have underlined the 
inefficiency of the Alfv\'{e}n wave scattering picture to capture
the observed properties of galactic cosmic rays \citep[e.g.,][]{chandran2000,yan_lazarian2004,chan2019,hopkins2022,Butsky2024}. Recently, \citet{Lemoine2023} and \citet{Kempski23} proposed mechanisms for strong CR scattering from non volume-filling, intermittent structures with a large magnetic field curvature, which in our description constitute MCoSs.
Along the same lines, \citet{Butsky2024} proposed a patchy scattering scenario for galactic CRs, and made order-of-magnitude estimates for the required size and volume-filling factor of the scattering structures. They concluded that strong scattering of CRs by intermittent regions is a very promising scenario for explaining the observed CR properties. 
Our work provides the statistics of these locations in global galaxy models.

Our statistical analysis of the MCoSs, which in the case of CR propagation would be scattering structures, shows that they are not uniformly distributed inside our galaxy as was assumed by several studies \citep[e.g.,][]{Lazarian23, Lemoine2023, Kempski23, Butsky2024}. Instead, they have a fractal structure. Transport and acceleration inside a fractal environment  
is a well-studied topic in space physics \citep[e.g.][]{Isliker03, Vlahos04, Sioulas20a, Sioulas2020b, Sioulas22} and has been shown to change the transport properties of the particles \citep{Vlahos08, Isliker2017}. Fundamentally, if the scattering structures are not volume-filling, the mean free path of the CR ($\lambda_{sc}$)  traveling inside the galaxy is not constant.  Also, particles transported inside fractal MCoSs are re-accelerated, rendering the mean free path energy-dependent: $\lambda_{sc}(E)$. This energy dependence adds a significant complication in studying their transport properties \citep{Bouchet04}.
The statistical description of MCoSs in this work provides a framework for studying these processes in the ISM of galaxies.

\section{Conclusions}
\label{sec:concl}

Our analysis reveals the presence of strong magnetic turbulence in the ISM of simulated galaxies, irrespective of the initial magnetic field morphology. In particular, after studying two galaxy-scale MHD simulations with initially fully ordered and initially fully stochastic
magnetic field, we conclude that:

\begin{itemize}
    \item The probability distribution of the magnetic field fluctuations, $\rm\delta B/B_0$, follows a log-normal distribution with a power-law tail at values $\rm\delta B/B_0>1$, for both models and since very early evolution times. This characteristic is indicative of dynamical processes like strong turbulence.
    \item Similarly, the current density distribution normalized over its rms value, $\rm J/J_{rms}$, also follows a log-normal probability distribution with a deviation at high values, which, however, is not power-law shaped. 
    \item Regions of strong current density, $\rm J/J_{rms}> \left(J/J_{rms}\right)_c$ 
    are embedded in regions of strong $\rm\delta B/B_0$, indicating an activation of MCoSs by strong magnetic turbulence.  
    \item Somewhat surprisingly, the regions of strongest current density coincide with regions of high star formation rate surface density. This connection could indicate that gravitational collapse motions drive abrupt changes in the magnetic field morphology.
    \item The fractal dimension of the MCoSs is close to unity a scales below 1~kpc and between 2 and 3 on larger scales, which is consistent with sheet-like regions containing filamentary dissipative structures. 
\end{itemize}

In the context of ISM dynamics, this first statistical description of MCoS properties can serve as a benchmark for determining the environment in which galactic CRs propagate.

\begin{acknowledgements}
    EN acknowledges funding from the Italian Ministry for Universities and Research (MUR) through the "Young Researchers" funding call (Project MSCA 000074).
    EN also thanks the KITP program "Turbulence in Astrophysical Environments", where part of this paper was written (grant NSF PHY-2309135 to the Kavli Institute for Theoretical Physics), for the generous hospitality and lively discussions. 
\end{acknowledgements}

\bibliographystyle{aa}
\bibliography{coh_structures}
\label{lastpage}

\clearpage

\appendix

\section{Time evolution of magnetic fluctuation and current density distributions}
\label{app:J_Jrms}

\begin{figure*}[h]
\centering
\includegraphics[trim={0.5cm 4.5cm 2cm 0cm},clip,width=0.7\textwidth]{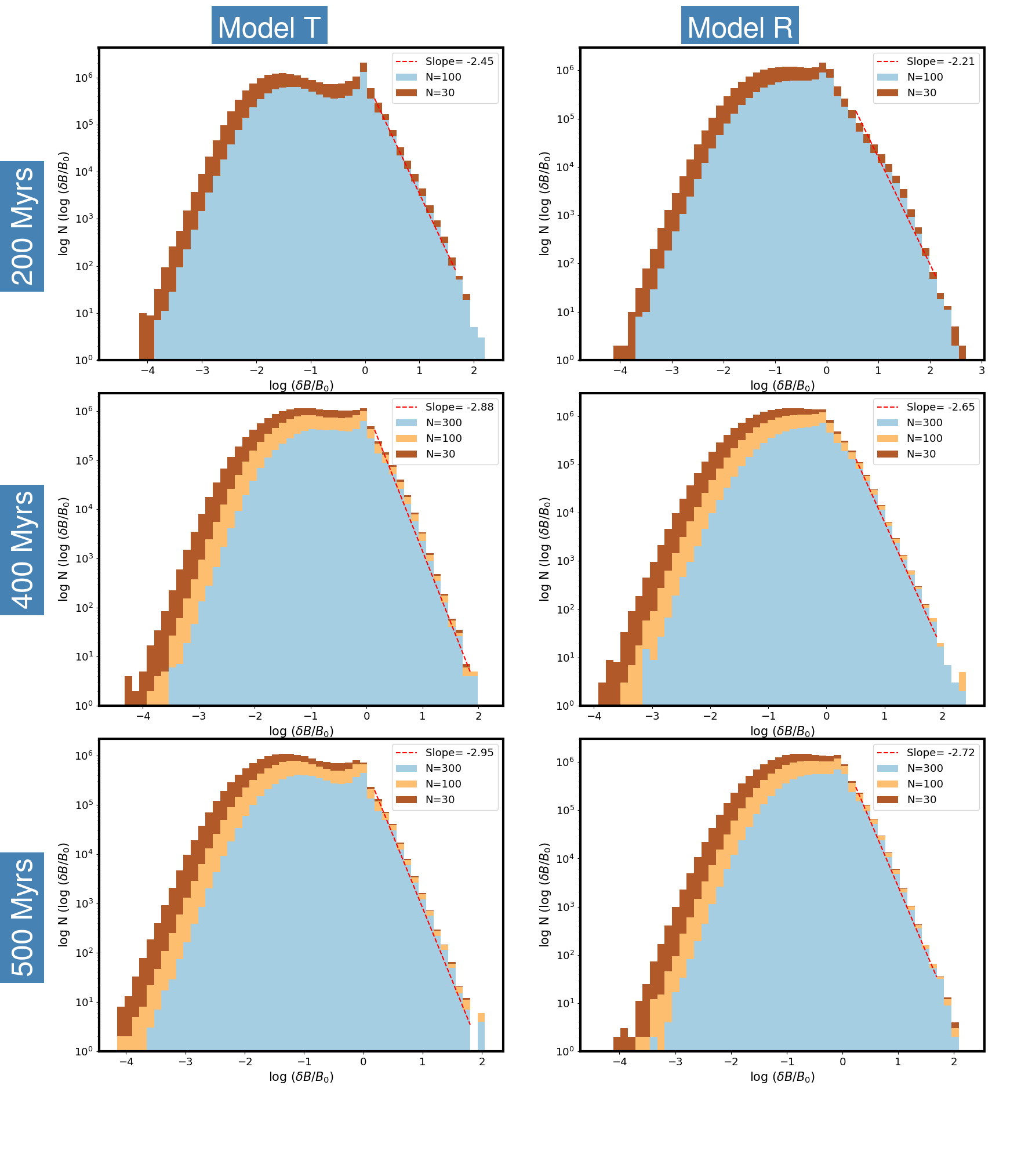}
\caption{Time evolution of the $\rm log~\delta B/B_0$ PDFs. From top to bottom, 200, 300 and 400~Myrs into the evolution of the galaxy models. The early snapshots have fewer refined cells, so the maximum number of neighbors used to calculate $B_0$ is limited to 100.}
\label{fig:dbb0_time}
\end{figure*}

Figure \ref{fig:dbb0_time} shows the time evolution of the $\rm log~\delta B/B_0$ PDFs. The early snapshots (t=200~Myrs) contain two levels of adaptive mesh refinement less than the later snapshots, therefore using N=300 is not practical, as it encompasses too large volumes.
The high-value ends of the PDFs have been fit by a power-law for the N=100 case, although the slope is identical for the other cases of N. 
From this sequence it is easy to appreciate the early onset of the power-law at high 
 $\rm log~\delta B/B_0$ values, as well as the consistent similarity of the PDFs between the two models, despite their very different initial conditions for the magnetic field.

\begin{figure*}
\centering
\includegraphics[trim={0.5cm 4.5cm 2cm 0cm},clip,width=0.7\textwidth]{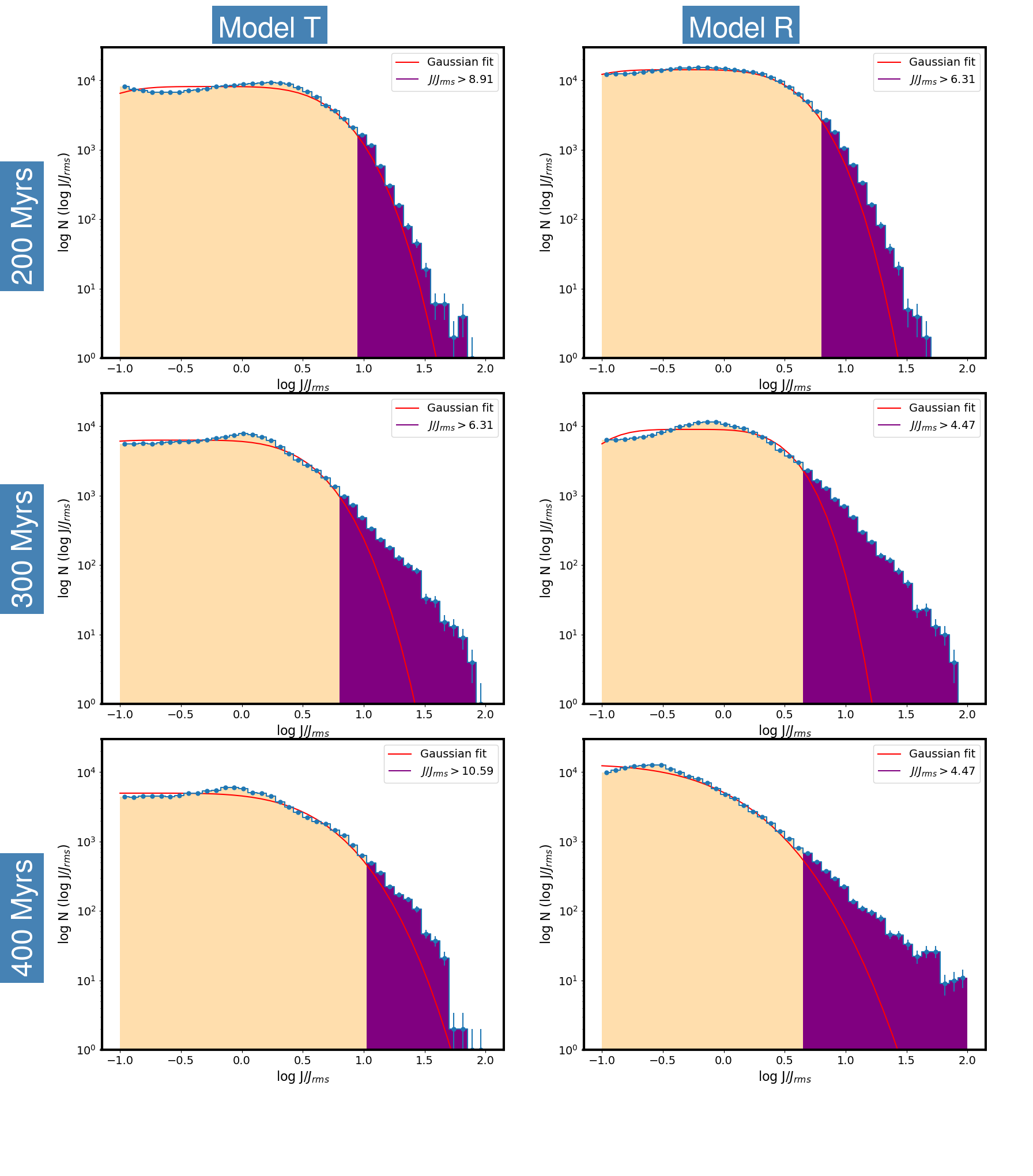}
\caption{Time evolution of the $\rm log~J/J_{rms}$ PDFs. From top to bottom, 200, 300 and 400~Myrs into the evolution of the galaxy models.}
\label{fig:J_Jrms_time}
\end{figure*}

Figure \ref{fig:J_Jrms_time} shows the time evolution of the $\rm log~J/J_{rms}$ PDFs at the high-value end (the full distributions expand to values tending to zero). We have fit a Gaussian to the distributions and marked the deviations from the fit in purple color, like in the main text. Unlike the $\rm \delta B/B_0$ PDFs, here the high-value end of the distribution is not power-law shaped, and it does change in time. There are some differences between the two models. The mean and variance of the Gaussian fit change with time and between the models, as does the limit for the Gaussian behavior. The origin of these variations is not clear from this comparison, and a wider parameter space for the numerical simulations would be needed in order to trace it.

\end{document}